\documentstyle[epsfig,pra,aps,twocolumn,floats]{revtex}
\draft
\begin{document}
\title{Direct sampling of 
exponential phase moments of smoothed Wigner functions
}
\author{Jarom\'{\i}r Fiur\'{a}\v{s}ek}
\address{Department of Chemical Physics, 
The Weizmann Institute of Science, Rehovot 76100, Israel}
\date{\today}
\maketitle
\draft

\begin{abstract}

We investigate exponential phase moments of the $s$-parametrized
quasidistributions (smoothed Wigner functions). We show that 
the knowledge of these moments as functions of $s$ provides, together
with photon-number statistics, a complete description of the
quantum state. We demonstrate that the exponential phase moments
can be directly sampled from the data recorded in balanced homodyne detection
and we present simple expressions for the sampling 
kernels. The phase moments are  Fourier coefficients
of phase distributions obtained from the quasidistributions via
integration over the radial variable in polar coordinates.
We performed Monte Carlo simulations of the homodyne detection
and we demonstrate the feasibility of  direct sampling of the moments
and subsequent reconstruction of the phase distribution.

\end{abstract}


\section{Introduction}

Quantum-state tomography is a powerful tool allowing us to reconstruct the
quantum state of a traveling optical mode, provided that many identical copies
of the state can be prepared \cite{Smithey93a,Breitenbach95}. 
The idea of homodyne tomography stimulated research in the field 
of quantum-state reconstruction of other simple quantum-mechanical systems.
Recently, reconstructions of the quantum state of a molecular vibrational mode
\cite{Dunn95} and the motional quantum state of a trapped ion
\cite{Wallentowitz95,Leibfried96} have been reported.

Optical homodyne tomography relies on balanced homodyne detection. 
The signal field is mixed with a strong coherent
local oscillator (LO) at a lossless $50/50$ beam splitter. 
Both the LO and the signal are derived from a common master
oscillator to ensure a stable phase difference $\theta$ between them. 
Two photodetectors are placed at the output ports of the beam splitter 
and the measured photocurrents are subtracted. The resulting signal is
proportional to the rotated  quadrature of the signal mode
$x_\theta$. The measurement, which yields the probability distribution
$w(x_\theta,\theta)$ of the quadrature $x_\theta$, is repeated for many
different phase  shifts $\theta$ from interval $[0,2\pi]$.

The Wigner function of the signal mode can be recovered from
the measured statistics $w(x_\theta,\theta)$ by means of inverse Radon
transform \cite{Vogel89,Smithey93a}. 
Numerical implementation of this inversion is not simple
and a filtering algorithm has to be applied to 
achieve the desired reconstruction. To avoid these complications, it was
suggested to directly get quantities of interest from the measured data
 by averaging appropriate kernels
over the distributions $w(x_\theta,\theta)$. This approach proved
to be very fruitful, and  kernels for the direct sampling of
density-matrix elements in the Fock basis $\rho_{mn}$ \cite{Fock},
the moments $\langle a^{\dagger j}a^k\rangle$ \cite{Richter96},
Fourier coefficients of the canonical phase distribution \cite{Dakna98},
and for smoothed Wigner functions \cite{Richter99} have been found.
A different approach to the quantum-state reconstruction employs
a maximum likelihood estimation \cite{Hradil95}.
It was demonstrated recently that this technique 
can be used to estimate photon number distribution 
\cite{Banaszek98} and even a whole density matrix \cite{Banaszek99}.
For a review, see \cite{Welsch99}.

In recent years, great attention has been devoted to the quantum phase. 
Canonical phase distribution introduced by London \cite{London27} 
represents a limit of Pegg-Barnett phase formalism  \cite{Pegg88}. 
Recently, an approximate measurement of the canonical phase distribution,
using the phase-coherent states, has been proposed \cite{Paris99}.
One can also construct phase distributions from the  phase-space
quasidistributions  \cite{Schleich89,Braunstein90,Tanas96}. 
The phase distribution obtained from the $Q$ function (or smoothed $Q$ function
in the case of imperfect detection) can be directly
measured \cite{Schleich92,Leonhardt93b}.
An operational approach to the quantum phase, based on the 
description of a given experimental setup, has been proposed
by Noh {\em et al.} \cite{Noh91}.
The relation between canonical and measured phase distributions was discussed
in \cite{Leonhardt95b}. For a recent review, see \cite{Tanas96,Perinova98}.

Canonical phase distribution as well as phase distributions obtained
from quasidistributions cannot be directly sampled from the 
homodyne data. One has to reconstruct the Wigner function or the
density matrix and then use the definition of the
phase distribution to calculate it \cite{Beck93}. 
This detour via the Wigner function or the
density matrix complicates numerical data processing and 
increases error in the final result. 
However, the exponential phase moments 
(Fourier coefficients)  of the canonical phase distribution can 
be directly sampled with the use of appropriate kernels
\cite{Dakna98}. Phase-number uncertainty relations can be verified
by sampling the first exponential moment of the canonical phase distribution 
and the photon-number variance \cite{Opatrny98}.
It was also pointed out in \cite{Dakna98}
that the  exponential phase moments of the Wigner function 
can be directly sampled.

\vspace*{38mm}
\hspace*{10mm}
\copyright $~$ 2000 The American Physical Society
\vspace*{-42mm}
\hspace*{-10mm}

But we do not have to restrict ourselves to the exponential phase moments
of canonical phase distribution or the Wigner function.
In this paper, we consider direct sampling of the exponential
phase moments of general $s$-parametrized phase distributions. 
We show that it is possible to directly sample 
the exponential phase moments  of any $s$-parametrized quasidistribution
for $s<-(1-\eta)/\eta$, where $\eta$ is the overall detection efficiency.
Namely, we find the expressions for the kernels
whose average over data recorded in balanced homodyne detection
yields the exponential phase moments.
We show that a knowledge of these moments as functions of $s$ and the
photon-number distribution provides complete
characteristics of a given  quantum state.
The phase moments are Fourier coefficients of the phase distributions
defined  as radial integrals of the $s$-parametrized quasidistributions
in the polar coordinates. We demonstrate that these phase distributions 
can be successfully reconstructed from the sampled phase moments.

The paper is organized as follows. In Sec. II
the exponential phase moments are introduced and discussed.
In Sec. III simple analytical expressions
for the sampling kernels are derived and the influence of imperfect detection
is addressed. In Sec. IV the results of  Monte Carlo
simulations are presented. Section V contains conclusions. 
Some mathematical issues are linked to the Appendix.

\section{Exponential phase moments} 

The quasidistributions related to various $s$ orderings of creation
and annihilation operators can be expressed in terms of the density
matrix $\rho$ \cite{PBook},
\begin{equation}
W(\alpha,s)=\frac{1}{\pi^2}\int e^{s|\beta|^2 / 2}
{\rm Tr}\,\left[\rho e^{(a^{\dagger}-\alpha^{\ast})\beta-(a-\alpha)\beta^{\ast}}
\right]
d^2\beta,
\end{equation}
where $a$,$a^{\dagger}$ are annihilation and creation operators.
One gets the $P$ representation for $s=1$, the Wigner function for $s=0$
and the $Q$ function for $s=-1$.
The $s$-parametrized quasidistributions are mutually related
through the convolution
\begin{eqnarray}
W(q,p,s_2)&=&\frac{1}{\pi(s_1-s_2)}
\nonumber \\[1.8mm]
&&\times \int_{-\infty}^{\infty}
            \int_{-\infty}^{\infty}
    \exp\left[-\frac{(q-q^\prime)^2+(p-p^\prime)^2}{s_1-s_2} \right]
    \nonumber \\[1.8mm]
&&\times  W(q^{\prime},p^{\prime},s_1)\, dq^{\prime}\, dp^{\prime},
\label{SMOOTHW}
\end{eqnarray}
where $q=(\alpha+\alpha^{\ast})/\sqrt{2}$ and 
$p=-i(\alpha-\alpha^{\ast})/\sqrt{2}$ are the usual quadratures, 
and $s_1>s_2$ must hold.

It is convenient to introduce  polar coordinates
$q=r\cos\theta$, $p=r\sin\theta$.
The phase distribution  $P_s(\theta)$ related to $s$-parametrized
quasidistribution is defined as \cite{Tanas96}
\begin{equation}
P_s(\theta)=\int_0^{\infty}W(r,\theta,s)\, r \, d r.
\end{equation}
It should be noted that the phase distributions $P_s(\theta)$
can be negative for $s>-1$. 
Only the phase distributions obtained from the
$Q$ function (or the smoothed $Q$ function) are positively defined for every
quantum state. Moreover, for $s>0$, the distributions can be highly
singular generalized functions. Thus we restrict ourselves to
the negative $s$ in the following.

\begin{figure}[t]
\centerline{\epsfig{figure=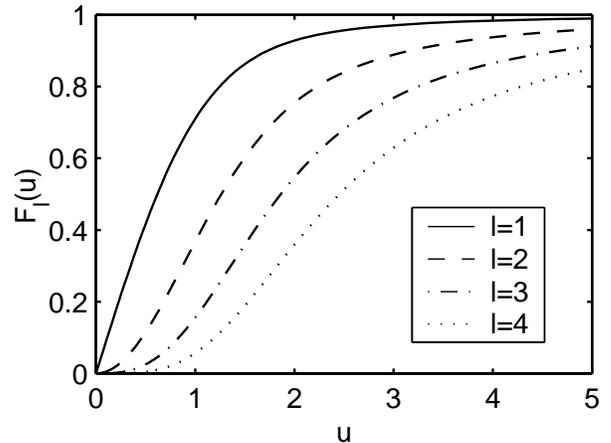,width=0.9\linewidth}}
\vspace*{3.5mm}
\caption{Filtering functions $F_l(u)$ for determination of 
exponential phase moments $\Psi_l(s)$ from the 
$s_0$-parametrized quasidistributions. }
\end{figure}

The exponential phase moments are defined as
\begin{eqnarray}
\Psi_l(s)&=&\langle\exp(il\theta)\rangle_s=
\int P_s(\theta)e^{il\theta}d\theta
\nonumber \\
&=&\int_{0}^{2\pi}\int_0^{\infty}W(r,\theta,s)e^{il\theta}\,r \,dr \,d\theta.
\label{PSIL}
\end{eqnarray}
The moments $\Psi_l(s)$ can be simply determined from any quasidistribution 
$W(q,p,s_0)$ provided that $s_0>s$. Indeed, 
inserting the relation (\ref{SMOOTHW}) into Eq. (\ref{PSIL}), we have
\begin{equation}
\Psi_l(s)=\int_{0}^{\infty}\int_{0}^{2\pi}
F_l\left(\frac{r}{\sqrt{s_0-s}}\right)
W(r,\theta,s_0) e^{il\theta}\, r \, d r \, d\theta.
\label{PSIWIGNER}
\end{equation}
The filtering functions  $F_l(u)$ are given by
\begin{equation}
F_l(u)=\frac{1}{\pi}\int_0^{\infty}\int_0^{2\pi}
e^{il\phi} e^{-u^2-\rho^2+2u\rho\cos\phi}
\rho \, d \rho \, d \phi.
\end{equation}
Integration over the angle variable $\phi$ yields  the modified Bessel
function $I_{l}(2u\rho)$. The resulting integral over radial 
variable $\rho$ can be found in the tables of integrals
(Ref. \cite{Prudnikov98}, p. 306, Eq. 2.15.5.4)  and we have
\begin{equation}
F_l(u)=\sqrt{\pi}\frac{u}{2}
\exp\left(-\frac{u^2}{2}\right)
 \left[I_{\frac{|l|-1}{2}}\left(\frac{u^2}{2}\right)
+I_{\frac{|l|+1}{2}}\left(\frac{u^2}{2}\right)\right].
\end{equation}
The first four filtering functions are plotted in Fig. 1.
They start from zero and asymptotically reach unity. 
The interval, where the functions $F_l(r/\sqrt{s_0-s})$ are significantly
lower than $1$, increases with decreasing $s$. This implies that
the absolute values of the  phase moments $\Psi_l(s)$  
decrease with decreasing $s$ because the modulation of the
phase distribution  $P_s(\theta)$ is suppressed by the smoothing 
convolution (\ref{SMOOTHW}).

It is remarkable that the functions 
$F_{l}(u)$ are closely related to the exponential phase moments of 
the coherent state $|\xi\rangle$,
\begin{equation}
\Psi_l(\xi;s)=F_{l}\left(\sqrt{\frac{2}{1-s}}\,|\xi|\right)e^{il\psi},
\qquad \psi=\arg\xi.
\label{FCS}
\end{equation}
To prove this, we notice that the quasidistributions $W_s(\alpha)$
of the coherent state $|\xi\rangle$ are shifted Gaussians,
\begin{equation}
W(\alpha,s)=\frac{2}{(1-s)\pi}\exp\left(-\frac{2|\alpha-\xi|^2}{1-s}\right).
\end{equation}
Inserting this into  Eq. (\ref{PSIL}), we immediately obtain Eq. (\ref{FCS}).

The filtering functions $F_l(u)$ can be expanded in Taylor series,
\begin{equation}
F_l(u)=\sum_{n=0}^{\infty}f_{n,l}\, u^{2n+|l|},
\label{FTAYLOR}          
\end{equation}
where
\begin{equation}
f_{n,l}=\frac{|l|}{2}(-1)^n\frac{\Gamma(n+|l|/2)}{n!\,(n+|l|)!}.
\end{equation}
It is convenient to introduce the parameter $t$, $s_0-s=1/t^2$.
With the help of the expansion (\ref{FTAYLOR}), we can rewrite
Eq. (\ref{PSIWIGNER}) as
\begin{eqnarray}
&&\Psi_{l}\left(s_0-\frac{1}{t^2}\right)=
\nonumber \\
&&\quad\sum_{n=0}^{\infty}f_{n,l}\int_{0}^{\infty}\int_{0}^{2\pi}
(rt)^{2n+|l|}\,W(r,\theta,s_0) e^{il\theta}\, r \, d r \, d\theta.
\label{PSITAYLOR}
\end{eqnarray}
It follows from this formula that $\Psi_l(s)$ are generating
functions of the $s_0$-ordered moments,
\begin{equation}
\langle r^{2n+|l|}e^{il\theta}\rangle_{s_0}=
\left.\frac{1}{(2n+|l|)! \, f_{n,l}}
\frac{d^{2n+|l|}}{d t^{2n +|l|}}
\Psi_l\left(s_0-\frac{1}{t^2}\right)\right|_{t=0}.
\label{PSIGENERATING}
\end{equation}
The limit $t\rightarrow 0$ should be taken only after 
the derivative is performed.  The generating functions
$\Psi_l(s)$ can be used to determine the moments 
$\langle r^{2n+|l|}e^{il\theta}\rangle_{s_0}$ for {\em any} ordering
parameter $s_0$.
Notice, however, that the formula (\ref{PSIGENERATING}) 
fails for $l=0$. The exponential
phase moments do not allow us to determine the moments $\langle r^{2n}\rangle$
which are related to photon-number statistics.
As an example, consider the Fock state $|n\rangle$. This state
is phase insensitive, $\Psi_l(s)=0$ for $l\neq 0$,
and the phase is uniformly distributed over the $2\pi$ interval,
$P_s(\theta)=1/2\pi$. 
Note also that the photon-number distribution $p(n)$
can be recovered from the phase-averaged quadrature distributions
\cite{Munroe95}.

The $s$-ordered moments (\ref{PSIGENERATING}) are simply related to more
familiar moments of creation and annihilation operators. 
With the help of $\alpha=2^{-1/2}r\exp(i\theta)$ we find that 
\begin{equation}
\langle a^{\dagger n}a^{n+l}\rangle_s=
2^{-(n+l/2)}
\langle r^{2n+l}e^{il\theta} \rangle_{s}
\end{equation}
and a similar expression holds for $\langle a^{\dagger n+l}a^{n}\rangle_s$.
The formula (\ref{PSIGENERATING}) allows us to find any moments
$\langle a^{\dagger m} a^{n}\rangle$ provided that $m\neq n$.
Complementarily, the moments
$\langle a^{\dagger k}a^k\rangle=\langle :\! n^{k} \! \!:\,\rangle$ 
can be determined from the photon-number distribution.

The phase moments $\Psi_l(s)$ are linear combinations of
density-matrix elements $\rho_{n+l,n}$,
\begin{equation}
\Psi_l(s)=\sum_{n=0}^{\infty}c_{n,l}(s)\rho_{n+l,n},
\label{PSIRO}
\end{equation}
where \cite{Tanas93}
\begin{eqnarray}
c_{n,l}(s)&=&\left(\frac{2}{1-s}\right)^{n+l/2} [n!\,(n+l)!]^{1/2}
\nonumber \\
&&\times \sum_{k=0}^n
\frac{\Gamma(n-k+l/2+1)}{k!\,(n-k)!\,(n+l-k)!}
\left(-\frac{1+s}{2}\right)^k.
\end{eqnarray}
If $l\neq 0$, the relation (\ref{PSIRO}) 
can be inverted and $\rho_{n+l,n}$ can be found
from $\Psi_l(s)$.  In principle, the knowledge of $\Psi_{l}(s)$ 
at an infinite but countable number of points $s_j$ can be sufficient
for determination of all $\rho_{n+l,n}$ from Eq. (\ref{PSIRO}).
Diagonal matrix elements appear only in
\begin{equation}
\Psi_0(s)=\sum_{n=0}^{\infty}\rho_{nn}\equiv{\rm Tr} \, \rho=1,
\end{equation}
and this relation cannot be inverted. Only when we know both
the phase moments $\Psi_{l}(s)$ and the photon-number distribution
$p(n)=\rho_{nn}$ can we determine all density-matrix elements
$\rho_{mn}$ or, equivalently,
 all moments $\langle a^{\dagger n}a^m\rangle_s$.
Thus the simultaneous knowledge of the functions $\Psi_l(s)$ and $p(n)$
provides complete information on the quantum state and it is
equivalent to the knowledge of the Wigner function or the density matrix.

\section{Sampling kernels for the exponential phase moments} 

Balanced homodyne detection provides statistics $w(x_\theta,\theta)$
of rotated quadratures,
\begin{equation}
x_\theta=\frac{1}{\sqrt{2}}\left(a e^{-i\theta}+a^{\dagger}e^{i\theta}\right),
\end{equation}
where  $\theta$ is the relative phase between the LO and the signal mode.
The probability distribution $w(x_\theta,\theta)$
can be obtained from the Wigner function $W(q,p)$ as a marginal
distribution \cite{Vogel89},
\begin{eqnarray}
w(x_{\theta},\theta)&=&
\int_{-\infty}^{\infty} \int_{-\infty}^{\infty}
            \delta(x_{\theta}-q\cos\theta-p\sin\theta)
            \nonumber \\[1.5mm]
           && \times W(q,p) \, dq \, dp.
 \label{PX}
\end{eqnarray}
We would like to sample the moments $\Psi_l(s)$ directly from the homodyne data
$w(x_{\theta},\theta)$ with the use of the kernels
${\cal{K}}_l(x_\theta,\theta;s)$:
\begin{eqnarray}
\Psi_l(s)=\int_{-\infty}^{\infty}\int_0^{2\pi}{\cal{K}}_l(x_\theta,\theta;s)
w(x_\theta,\theta) \, d x_\theta\, d\theta.
\label{FKER}
\end{eqnarray}
The $\theta$ dependence of the kernels must be of the form $\exp(il\theta)$
\cite{Dakna98}. Thus we look for the kernels in the form 
\begin{eqnarray}
{\cal{K}}_l(x_{\theta},\theta;s)=K_l(x_\theta,s)e^{il\theta}.
\end{eqnarray}
In what follows we restrict ourselves to positive $l$. For negative $l$,
the exponential moments can be obtained by complex conjugation,
$\Psi_{-l}(s)=\Psi_l^{\ast}(s)$.
Now we substitute Eq. (\ref{PX}) into Eq. (\ref{FKER}),
perform integration over $x_\theta$,
and rewrite the remaining integral in polar coordinates.
After some algebra, we arrive at
\begin{eqnarray}
\Psi_l(s)&=&
\int_{0}^{2\pi}\int_{0}^{\infty} 
\left[\int_{0}^{2\pi}
K_l(r\cos\phi,s)e^{il\phi}d\phi\right]
\nonumber \\[1.5mm]
&&\times W(r,\theta) e^{il\theta} r \, dr \, d\theta.
\label{PSI2}
\end{eqnarray}
\begin{figure}[t]
\centerline{\epsfig{figure=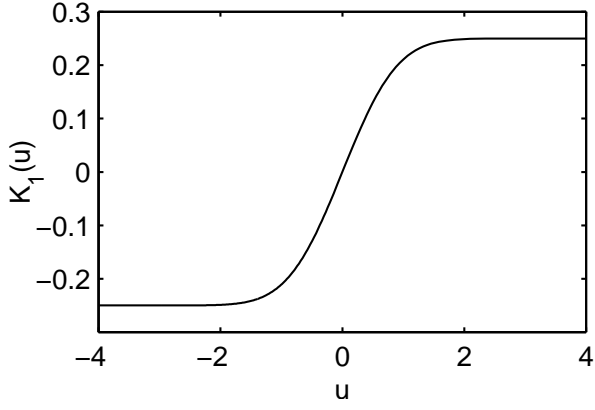,width=0.9\linewidth}}
\vspace*{3.5mm}
\caption{Kernel $K_1(u)$
for sampling of odd exponential phase moments.  }
\end{figure}
Comparing the formulas (\ref{PSI2}) and (\ref{PSIWIGNER}), where we set $s_0=0$,
we conclude that the kernel $K_l(x_\theta,s)$ must fulfill the integral equation
\begin{equation}
\int_{0}^{2\pi}
K_l(r\cos\theta,s)e^{il\theta}d\theta=F_l\left(r/|s|^{1/2}\right).
\label{FIE}
\end{equation}
In order to solve this equation,
we expand the kernel $K_l(x_\theta,s)$ in Taylor series,
\begin{equation}
K_l(x_{\theta},s)=\sum_{n=0}^{\infty}a_n(l,s) \, x_{\theta}^n.
\label{KTS}
\end{equation}
This expansion is inserted into Eq. (\ref{FIE}) and the integration 
over $\theta$ is carried out, using the formula
\begin{eqnarray}
\int_0^{2\pi} (\cos\theta)^{2n+l}e^{il\theta} d\theta=
\frac{2\pi}{2^{2n+l}}
{ 2n+l \choose n }.
\end{eqnarray}
Comparing the Taylor series on the left-hand side of Eq. (\ref{FIE})
with the series (\ref{FTAYLOR}), we find the coefficients
$a_n(l,s)$. Inserting them back into the series (\ref{KTS}),
we arrive at 
\begin{eqnarray}
K_l(x_\theta,s)=\frac{l}{4\pi}\sum_{n=0}^{\infty}(-1)^n
\frac{\Gamma(n+l/2)}{(2n+l)!}
\left(\frac{2x_\theta}{|s|^{1/2}}\right)^{2n+l}.
\end{eqnarray}


\begin{figure}[t]
\centerline{\epsfig{figure=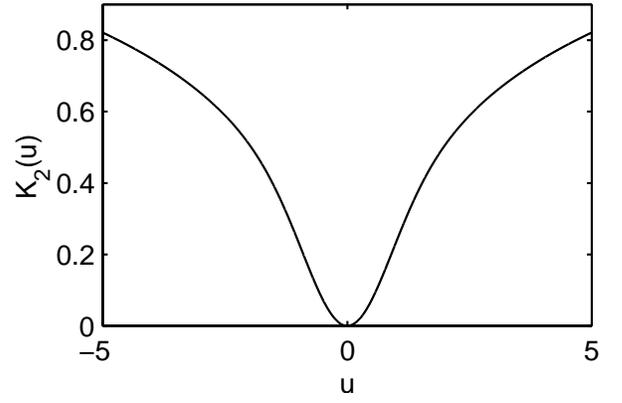,width=0.9\linewidth}}
\vspace*{3.5mm}
\caption{Kernel $K_2(u)$ 
for sampling of even exponential phase moments.  }
\end{figure}

Notice that the kernel is a function of a specific combination of $x_\theta$
and $s$, $u=x_\theta/|s|^{1/2}$.
In the following we use $u$ for simplicity.
Let us  discuss the relation between the kernels $K_{l}(u)$
and $K_{l+2}(u)$. We have
\begin{eqnarray}
K_{l+2}(u)&=&
-\frac{l+2}{4\pi}\sum_{n=1}^{\infty}(-1)^n
\frac{\Gamma(n+l/2)}{(2n+l)!}
\left(2u\right)^{2n+l}
\nonumber \\
&=&-\frac{l+2}{l}K_l(u)+\frac{l+2}{4\pi}\frac{\Gamma(l/2)}{l!}
   \left(2u\right)^l.
   \label{KERNELREL}
\end{eqnarray}
However, the kernels $K_l$ are not uniquely determined. 
Any polynomial of order lower than
$l$ can be added to kernel $K_l$, because all  such polynomials
 are solutions of the homogeneous 
integral equation 
\begin{equation}
\int_{0}^{2\pi}f(r\cos\theta)e^{il\theta}=0.
\end{equation}
Thus we can neglect
the last term in the formula (\ref{KERNELREL}) 
and we can define the kernels for which
\begin{eqnarray}
K_{l+2}(u)=-\frac{l+2}{l}K_{l}(u)
\label{REC}
\end{eqnarray}
holds.
It remains to find out the kernels $K_1$ and $K_2$. 
The summation of the series can be found in the Appendix. 
The results are
\begin{eqnarray}
K_{1}(u)=\frac{1}{4}\,{\rm erf}\left( u\right),
\end{eqnarray}
\begin{eqnarray}
K_2(u)=\frac{1}{\sqrt{\pi}}
         \int_{0}^{u}e^{-y^2}\,{\rm erfi}(y)\,dy.
\end{eqnarray}
The kernels are plotted in Fig. 2 and Fig. 3, respectively.

Combining this result with the recurrence formula (\ref{REC}),
we finally have
\begin{eqnarray}
{\cal{K}}_{2l+1}(x_\theta,\theta;s)&=&(-1)^l(2l+1)
K_1\left(x_\theta/|s|^{\frac{1}{2}}\right)e^{i(2l+1)\theta},
\nonumber \\
{\cal{K}}_{2l}(x_\theta,\theta;s)&=&
(-1)^{l-1} l K_2\left(x_\theta/|s|^{\frac{1}{2}}\right)e^{i2l\theta}.
\label{KERNELSFIN}         
\end{eqnarray}
For large $x_{\theta}$, all the kernels  tend to the same limit
because we move to the  strong classical field domain 
and the differences between various $s$ orderings vanish.
The limit for odd kernels is straightforward. We simply notice that
\begin{equation}
\lim_{x\rightarrow \pm\infty} {\rm erf}(x)=\pm 1   .
\end{equation}
The limit for even kernels can be found if we take into account that
for large $x$,
\begin{equation}
e^{-x^2}{\rm erfi}(x)\approx\frac{1}{\sqrt{\pi}}\frac{1}{x}.
\end{equation}
Inserting this into Eq. (\ref{KERNELSFIN}) we have for large $x_{\theta}$
\begin{equation}
{\cal{K}}_{2l}(x_\theta,\theta;s)\approx
\frac{1}{\pi}l(-1)^{l-1} \ln|x_{\theta}|e^{i2l\theta} +C_{l,s}e^{i2l\theta} .
\end{equation}
Here $C_{l,s}$ is some constant. The superfluous term containing this constant 
can be omitted for reasons discussed above and we can see that
as a limit all kernels approach those for the 
phase moments of the Wigner function \cite{Dakna98}:
\begin{eqnarray}
{\cal{K}}_{2l+1}&=&\frac{1}{4}(2l+1)(-1)^{l}\,{\rm sgn}(x_{\theta})\,
e^{i(2l+1)\theta},
\nonumber \\
{\cal{K}}_{2l}&=&\frac{1}{\pi}l(-1)^{l-1}\,{\ln}|x_{\theta}|\,e^{i2l\theta}.
\label{KWIGNER}
\end{eqnarray}

Up to now, we have considered ideal detectors having unit 
quantum efficiency. In a realistic experiment, the detection
efficiency $\eta$ is lower than $100\%$ and 
the smoothed quadrature distributions
$w(x_\theta,\theta;\eta)$ are recorded \cite{Vogel93},
\begin{eqnarray}
w(x_\theta,\theta;\eta)&=&\frac{1}{\sqrt{\pi(1-\eta)}}
\nonumber \\[1.7mm]
&&\times\int_{-\infty}^{\infty} w(x_{\theta}^{\prime},\theta)
\exp
\left[-\frac{(x_{\theta}-\sqrt{\eta}x_{\theta}^{\prime})^2}{1-\eta}\right]
\,d\, x_\theta^{\prime}.
\nonumber \\
\end{eqnarray}
The smoothed quadrature distributions $w(x_\theta,\theta;\eta)$
can be obtained from the scaled and smoothed Wigner function,
\begin{eqnarray}
w(x_\theta,\theta;\eta)&=&
\int_{-\infty}^{\infty}\int_{-\infty}^{\infty}
\delta(x_{\theta}-q\cos\theta-p\sin\theta)
\nonumber \\[1.7mm]
&&\times \frac{1}{\eta}
W\left(\frac{q}{\sqrt{\eta}},\frac{p}{\sqrt{\eta}},-\frac{1-\eta}{\eta}\right)
\,d q \, d p.
\end{eqnarray}
The scaling and smoothing are two factors which must be included in
the kernels ${\cal{K}}_{l}(x_\theta,\theta; s , \eta)$.
The scaling means that we must
replace $x_\theta$ by $x_\theta/\sqrt{\eta}$. The smoothing tells us
that the kernels ${\cal{K}}(x_\theta,\theta,s)$ would provide us
with exponential phase moments $\Psi_l(s+s_\eta)$, $s_\eta=-(1-\eta)/\eta$.
Thus we must replace $s$ with $s-s_\eta$ in all
the expressions (\ref{KERNELSFIN}). 
It is obvious that the losses impose a new limit.
We can reconstruct only exponential phase moments for the phase distributions
corresponding to $s<s_\eta$. The modified kernels are
\begin{equation}
{\cal{K}}_l(x_\theta,\theta; s, \eta)=
{\cal{K}}_l\left(x_\theta/\sqrt{\eta},\theta;s+(1-\eta)/\eta\right),
\label{LOSSK}
\end{equation}
and the condition $s<-(1-\eta)/\eta$ must be fulfilled.

\begin{figure}[t]
\centerline{\epsfig{figure=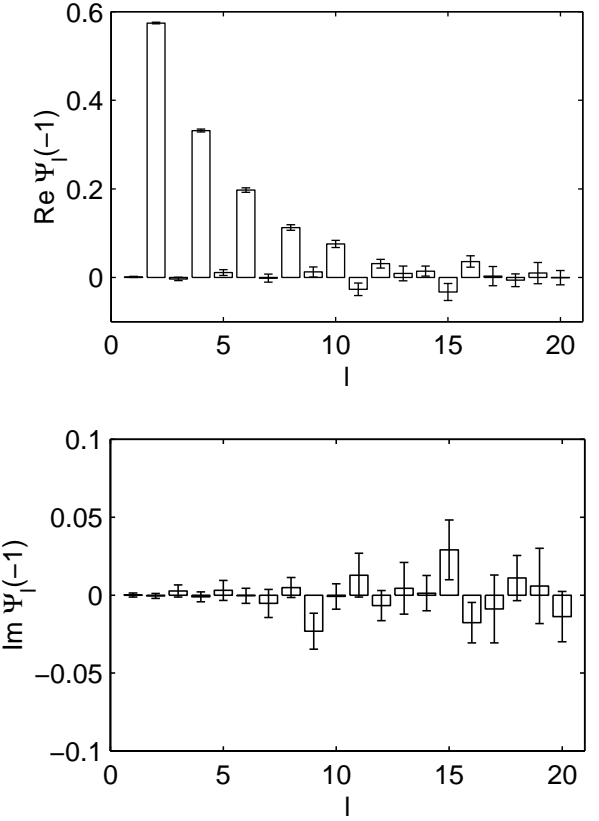,width=0.9\linewidth}}
\vspace*{3.5mm}
\caption{ Reconstructed phase moments $\Psi_{l}(-1)$
of squeezed vacuum state $|\zeta\rangle$, $\zeta=1.317$, i.e. 
$\langle n\rangle=3$. Statistical errors are denoted by error bars.
}
\end{figure}

\section{Monte Carlo simulations}

In order to test the kernels, we performed  Monte Carlo simulations
of the homodyne detection and  we present here the results of 
simulations for the squeezed vacuum state $|\zeta\rangle$,
\begin{equation}
|\zeta\rangle=\exp\left(\frac{1}{2}\zeta a^{\dagger 2}
-\frac{1}{2}\zeta^{\ast}a^2\right)|0\rangle,
\end{equation}
where $|0\rangle$ is the vacuum state.
The squeezed vacuum state belongs to the class of Gaussian states, i.e. states
whose quasidistributions $W(q,p,s)$ have Gaussian form. 
The phase distribution $P_s(\theta)$ for the general Gaussian mixed state
was determined in \cite{Tanas93,Mista98}. 
In particular, it holds that $P_s(\theta)$ of the squeezed vacuum state
can be expressed as \cite{Mista98}
\begin{equation}
P_s(\theta)=\frac{1}{2\pi}\frac{(B_s^2-C^2)^{1/2}}{B_s-C\cos(2\theta-\psi)},
\end{equation}
where
\begin{eqnarray}
B_s&=&\sinh^2|\zeta|+(1-s)/2,
\nonumber \\
C&=&\frac{1}{2}\sinh(2|\zeta|),
\end{eqnarray}
and  $\psi=\arg \zeta$. The  phase moments 
can be calculated with the help of the residue theorem. One arrives at
\begin{eqnarray}
\Psi_{2l}(s)&=&\left(B_s/C-\sqrt{B_s^2/C^2-1}\right)^le^{il\psi},
\nonumber \\
\Psi_{2l-1}(s)&=&0.
\label{PSISVS}
\end{eqnarray} 
In our simulations, the sampling was performed for 120 values of $\theta$ 
equidistantly placed at the interval $[0,2\pi]$ and  5000 samples
have been made for each $\theta$.  We assumed that the overall detection 
efficiency is $\eta=80\%$ and we used the loss compensating 
kernels (\ref{LOSSK}).

\begin{figure}
\centerline{\epsfig{figure=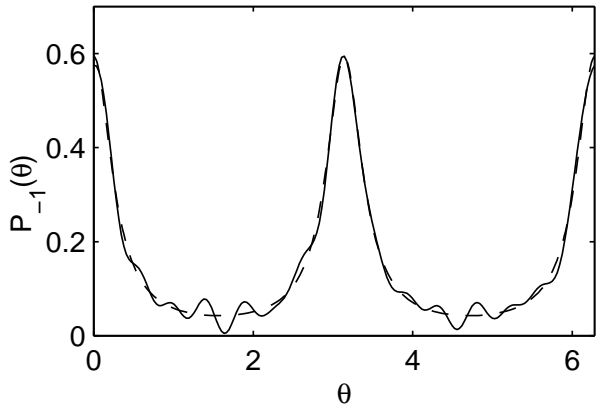,width=0.9\linewidth}}
\vspace*{3.5mm}
\caption{Reconstruction of the phase distribution $P_{-1}(\theta)$
of the squeezed vacuum state from the sampled 
exponential moments depicted in Fig. 4.
The solid line shows the reconstructed distribution and  the dashed line
represents the exact shape.}
\end{figure}

Figure 4 shows the reconstructed phase moments of the
$Q$ function, $\Psi_{l}(-1)$.
The results are in very good agreement with the exact values following from
Eq. (\ref{PSISVS}). Statistical errors were calculated
in a manner described in \cite{Dakna98}.
As a rule,  error increases with increasing $l$ and this uncertainty
is responsible for the fast oscillations in the reconstructed probability
distribution $P_{-1}(\theta)$, see Fig. 5.

The reconstructed  moments $\Psi_l(s)$, considered as functions
of the ordering parameter $s$, are plotted in Fig. 6. Again, we found that
the curves are in good agreement with their theoretical counterparts.
Notice that, due to the assumed $80\%$ efficiency of the detection,
we were able to sample only moments for $s<-0.25$.

We repeated our simulations also for other types of quantum states such as 
coherent states and displaced Fock states. In all cases, 
the reconstruction procedure worked well 
and the sampled moments were in good agreement with the theoretical
values. We emphasize that we have used only $6\times 10^5$ 
samples in our simulations and such an amount of data can be routinely
recorded in the experiment.

\begin{figure}
\centerline{\epsfig{figure=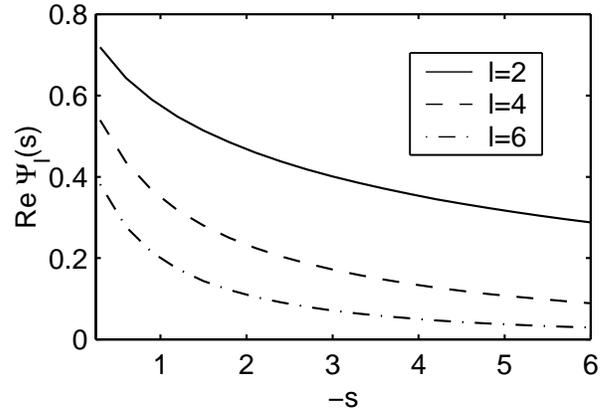,width=0.9\linewidth}}
\vspace*{3.5mm}
\caption{Reconstructed phase moments $\Psi_{l}(s)$
as functions of the $s$ parameter.
}
\end{figure}

\section{Conclusions}
We have shown that  the exponential phase moments
of the $s$-parametrized quasidistributions  are generating
functions of the moments of creation and annihilation operators. 
A simultaneous knowledge of photon-number distribution and
the functions $\Psi_l(s)$  provides a complete description of the
quantum state.
We have found kernels for direct sampling of  the  moments $\Psi_l(s)$  
from  quadrature distributions measured in optical homodyne detection.
The detection efficiency $\eta$ imposes a bound on the
ordering parameter, we can sample only phase moments for
$s<-(1-\eta)/\eta$. In the ideal case $\eta=1$ and the Wigner function
represents the limit; for $\eta=0.5$ the limit is formed by a $Q$ function.
We performed numerical Monte Carlo simulations
of homodyne detection, thereby
demonstrating the feasibility of direct sampling of the
exponential  phase moments from experimental data.

\acknowledgments

I would like to thank  T. Opatrn\'{y},
J. Pe\v{r}ina, and I.Sh. Averbukh for helpful discussions. 
I am pleased to acknowledge 
support of the U.S.-Israel Binational Science Foundation (Grant No. 96--00432).

\appendix
\section*{Summation of the series for the kernels $K_1$ and $K_2$}

Here we sum the Taylor series for kernels $K_1(u)$ and $K_2(u)$.
We start with $K_1(u)$. Using the formula for the
Gamma function of a half-integer,
\begin{equation}
\Gamma(n+1/2)=\sqrt{\pi}\,\frac{(2n)!}{2^{2n}n!},
\end{equation}
the series for $K_1(u)$ take on the form
\begin{equation}
K_1(u)=\frac{1}{4\pi}\sum_{n=0}^{\infty}(-1)^n
\frac{2\sqrt{\pi}}{n!}\frac{u^{2n+1}}{(2n+1)}.
\end{equation}
The derivative of the kernel is
\begin{equation}
\frac{d}{du}K_1(u)=\frac{1}{4}\sum_{n=0}^{\infty}(-1)^n
\frac{2}{\sqrt{\pi}}\frac{u^{2n}}{n!}
=\frac{1}{4}\frac{2}{\sqrt{\pi}} e^{-u^2}.
\end{equation}
Integrating the above equation we arrive at
\begin{equation}
K_1(u)=\frac{1}{4}\,{\rm erf}(u).
\end{equation}
We adopt a similar approach to determine $K_2(u)$,
\begin{equation}
K_2(u)=\frac{1}{2\pi}\sum_{n=0}^{\infty}(-1)^n
\frac{n!}{(2n+2)!}(2u)^{2n+2}.
\end{equation}
We calculate the derivatives
\begin{equation}
f(u)=\frac{d}{du}K_2(u)=
\frac{2}{2\pi}\sum_{n=0}^{\infty}(-1)^n
\frac{n!}{(2n+1)!}(2u)^{2n+1}
\end{equation}
and
\begin{eqnarray}
f^{\prime}(u)&=&\frac{d^2}{du^2}K_2(u)=
\frac{4}{2\pi}\sum_{n=0}^{\infty}(-1)^n
\frac{n!}{(2n)!}(2u)^{2n}
\nonumber \\
&=&\frac{2}{2\pi}\sum_{n=1}^{\infty}(-1)^n
\frac{(n-1)!}{(2n-1)!}(2u)^{2n}+\frac{2}{\pi}.
\end{eqnarray}
Thus we have
\begin{equation}
f^{\prime}(u)=-2uf(u)+\frac{2}{\pi}.
\end{equation}
Let us look for the function $f(u)$ in the form
\begin{equation}
f(u)=\frac{2}{\pi}\frac{g(u)}{g^{\prime}(u)}.
\label{FAPP}
\end{equation}
Substituting this into the above equation, we finish with
\begin{eqnarray}
g^{\prime\prime}(u)&=&2ug^{\prime}(u),
\nonumber \\
g(u)&=&\frac{\sqrt{\pi}}{2}\,{\rm erfi}(u).
\end{eqnarray}
Inserting this into Eq. (\ref{FAPP}), we conclude that
\begin{equation}
K_2(u)=\frac{1}{\sqrt{\pi}}\int_0^{u}e^{-y^2}{\rm erfi}(y)\, d y.
\end{equation}

\end{document}